# Epitaxial Growth of 2D Layered Transition Metal Dichalcogenides

Tanushree H. Choudhury,[1] Xiaotian Zhang,[2,3] Zakaria Y. Al Balushi,[4] Mikhail Chubarov[1] and Joan M. Redwing[1,2*]

[1]2D Crystal Consortium – Materials Innovation Platform
[2]Department of Materials Science and Engineering
The Pennsylvania State University, University Park, PA USA

[3]Advanced Light Source,
E.O. Lawrence Berkeley National Laboratory
Berkeley, CA 94720, USA

[4]Department of Materials Science and Engineering
University of California, Berkeley, CA USA

*Corresponding author: jmr31@psu.edu

**Keywords:** Transition metal dichalcogenide, $MoS_2$, $WSe_2$, epitaxy, vapor phase deposition

## Abstract

Transition metal dichalcogenide (TMD) monolayers and heterostructures have emerged as a compelling class of materials with transformative new science that may be harnessed for novel device technologies. These materials are commonly fabricated by exfoliation of flakes from bulk crystals, but wafer-scale epitaxy of single crystal films is required to advance the field. This article reviews the fundamental aspects of epitaxial growth of van der Waals bonded crystals specific to TMD films. The structural and electronic properties of TMD crystals are initially described along with sources and methods used for vapor phase deposition. Issues specific to TMD epitaxy are critically reviewed including substrate properties and film-substrate orientation and bonding. The current status of TMD epitaxy on different substrate types is discussed along with characterization techniques for large area epitaxial films. Future directions are proposed including developments in substrates, *in situ* and full wafer characterization techniques and heterostructure growth.

**E-mail and ORCID:**

T. H. Choudhury: tuc21@psu.edu; 0000-0002-0662-2594
X. Zhang: tiantian735@gmail.com; 0000-0002-1571-1257
Z. Y. Al Balushi: albalushi@berkeley.edu; 0000-0003-0589-1618
M. Chubarov: mikhail.chubarov@gmail.com; 0000-0002-4722-0321
J. M. Redwing: jmr31@psu.edu; 0000-0002-7906-452X

1. **Introduction**

Transition metal dichalcogenides (TMDs) form a compelling class of 2D materials with potential applications in optoelectronics, flexible electronics, chemical sensing and quantum technologies. At the monolayer limit, the semiconducting TMDs (*e.g.*, $MX_2$ where M = Mo/W and X = S/Se) exhibit direct band gaps within the visible range and large exciton binding energies (1, 2). The lack of out-of-plane bonding on the van der Waals surface of these materials enables heterostructure formation without the constraints of lattice matching (3). Furthermore, the lack of inversion symmetry in the monolayer results in unique optical selection rules, enabling circularly polarized light to selectively populate degenerate valleys in the Brillouin zone providing an additional degree-of-freedom for the realization of valleytronic devices (4). There is also growing interest in the properties of metallic TMDs such as $VS_2$ and $VSe_2$ which exhibit ferromagnetism and $1T-MoSe_2$ and $Td-WTe_2$ which are Weyl semimetals (5).

TMDs crystallize as layered compounds with metal atoms sandwiched between chalcogen atoms with stoichiometry of 1:2, respectively. The metal and chalcogens are covalently bonded within a layer and the layers are held together by van der Waals bonds (**Figure 1a**) (6). As a result of the weak interaction between layers, a single layer (referred to as a monolayer) or stack of a few layers can be readily exfoliated from the bulk crystal. Approximately 40 different TMD compounds exist which are comprised of transition metals from Groups 4, 5, 6, 7, 9 and 10 of the period table (**Figure 1b**) along with S, Se and Te.

The commercial availability of naturally occurring $MoS_2$ and chemically synthesized bulk crystals has made the field accessible to researchers for fundamental property studies and device development. Monolayer or few-layer TMD flakes can be readily exfoliated and manually stacked to produce van der Waals heterostructures without the need for sophisticated processing equipment. The size of exfoliated flakes, however, is typically limited to a few millimeters and properties may vary from flake to flake. Furthermore, residual water, organics and/or other impurities can be introduced at stacked interfaces unless processing occurs in a glove box or vacuum environment. As a result of these limitations, there is increasing interest in the epitaxial growth of wafer-scale single crystal TMD films and heterostructures with controllable layer number. This article provides an overview of issues pertinent to epitaxial growth of TMDs including a review of the current status of TMD epitaxy on both conventional and van der Waals-bonded substrates. Challenges and future directions for epitaxial materials synthesis are highlighted within the context of emerging research directions both for TMDs and the general field of van der Waals-bonded solids.

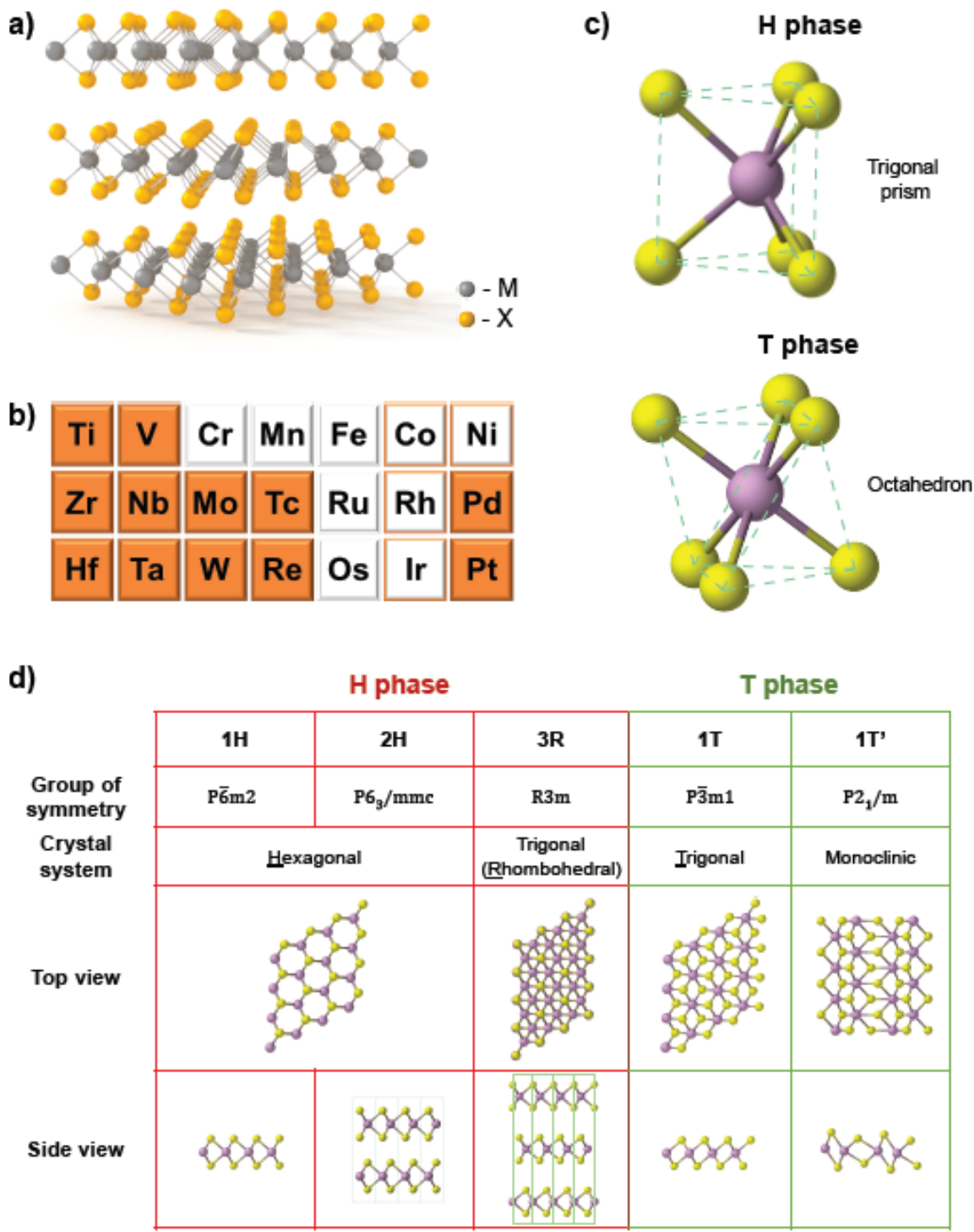

**Figure 1.** (a) Crystal structure of TMDs designated as $MX_2$ where M=transition metal and X=chalcogen. (b) Transition metals that crystallize as layered dichalcogenides (full orange). Co, Rh, Ir and Ni have dichalcogenide phases that are not layered (orange edge). (c) Primary phases for 1:2 (metal:chalcogen) TMD: H phase and T phase. (d) Crystal structures (top and side view) of common TMD phases.

## 1.1 Crystal structure and electronic properties of TMDs

The coordination of the metal and chalcogen atoms in a TMD can be viewed as consisting of two tetrahedrons that are stacked in opposite directions (**Figure 1c**) (5). The particular phases, which are designated as the H and T phases, are defined by the arrangement of the tetrahedra. In the H-phase, the upper and lower tetrahedra are arranged symmetrically to form a trigonal prismatic structure. In the T-phase, the upper tetrahedron is rotated by 180° forming an octahedral structure which is typically distorted.

The stacking sequence of layers, also referred to as the polytype, impacts the electronic structure, vibrational modes and optical properties. The polymorph or polytype is designated using a number in the phase nomenclature. For example, the monolayer form of the H phase consists of a hexagonal crystal system and therefore is named as the 1H phase (**Figure 1d**). The H phase can have varied polytypes due to different stacking sequence of layers. The 2H phase, which is the most common one, adds a screw rotation in the symmetry resulting from the AB stacking of the second layer while maintaining the hexagonal crystal system. The 3R phase reduces the rotation symmetry from 6-fold to 3-fold resulting in a rhombohedral lattice. On the other hand, the T phase of both monolayer and multilayer have the same symmetry owing to the AA stacking and therefore they can all be called 1T. However, the transition metals can also dimerize resulting in distortion of the chalcogen atoms out-of-plane which in turn changes the symmetry from 3-fold to 2-fold as exemplified by the 1T' phase. The distorted T phase (Td) is similar to the 1T' but exhibits a different c-axis angle.

The electronic properties of TMD bulk crystals are dependent on the coordination and number of d-electrons in the transition metal (7). For monolayer TMDs (1H and 1T phases), the non-bonding d-bands of the TMDs are in the gap between the bonding and antibonding bands of M-X bonds. The extent of filling of the non-bonding *d*-bands determines whether the material is semiconducting or metallic and also whether it exhibits magnetic properties. When the orbitals are partially filled, as is the case for Group 5 and Group 7, the TMD exhibits metallic conductivity as is the case for 1H-NbSe$_2$ and 1T-ReS$_2$. When the orbitals are fully occupied, as is the case for Groups 4, 6 and 10, the TMD is a semiconductor as is the case for 1T-HfS$_2$, 1H-MoS$_2$ and 1T-PtS$_2$. The chalcogen atom has less influence on the electronic properties compared to the transition metal, however, a general trend is toward a decreasing bandgap moving from S to Se to Te.

A reduction in the number of layers of the TMD results in a significant change in electronic properties which is particularly pronounced in semiconducting TMDs. Bulk 2H-MoS$_2$, for example, is an indirect-gap semiconductor while monolayer MoS$_2$ is direct gap at the K-point (**Figure 2a**) (8). Furthermore, as a result of the lack of inversion symmetry, there are two inequivalent momentum valleys, K and K' within the first Brillouin zone of monolayer MoS$_2$ (**Figure 2b**) (9). As a result of strong spin-orbit coupling, the valence band at these valleys splits with opposite spin splitting for K and K'. Electrons with a particular spin can be excited via circularly polarized light to populate one of the two valleys which is known as valley polarization. The control of valley polarization and its detection using light or electric fields offers an additional degree of freedom which may be exploited in the development of "valleytronic" switching devices.

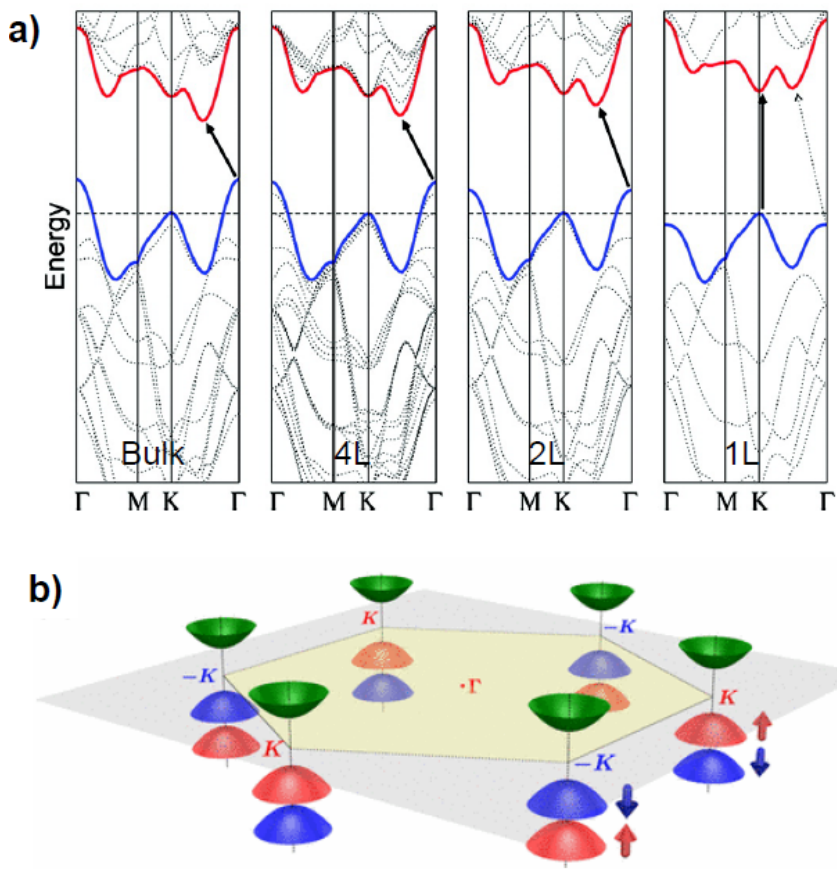

**Figure 2.** (a) Energy dispersion of bulk, 4 layer (4L), bilayer (2L) and monolayer (1L) $MoS_2$. Reprinted with permission from (8); (b) Band structure of $MoS_2$ showing inequivalent valleys K and -K (K') at the corners of the first Brillouin zone in $MoS_2$ and spin splitting in the valence band. Reprinted with permission from (9).

## 2. Techniques for Vapor Phase Deposition

Thin film epitaxy typically requires the use of vapor phase deposition methods such as molecular beam epitaxy (MBE) and chemical vapor deposition (CVD)/metalorganic CVD (MOCVD) which enable tight control of the source flux to the growth surface. The different techniques are distinguished by the source utilized (elemental vs. molecular precursor), the vapor phase transport process (molecular vs. viscous flow) and the role of precursor reaction chemistry. This section provides a brief overview of these topics specific to vapor phase deposition of TMDs.

## 2.1 Vapor phase sources

The transition metals and chalcogens, as pure elements, exhibit significantly different physical properties which have implications in the methods and conditions used for TMD synthesis. The transition metals have high bond strengths, high melting temperatures (>2000°C) and low elemental vapor pressures even at temperatures as high as 1000°C (**Table 1**). In contrast, the chalcogens have much lower melting temperatures (< 500°C) and correspondingly higher elemental vapor pressures. Sulfur, for example, has a vapor pressure greater than atmospheric pressure at temperatures above 500°C (10). These differences give rise to the characteristic phase diagrams of binary TMD compounds. In the case of the Mo-S phase diagram, for example, stoichiometric $MoS_2$ melts at 1730°C forming a Mo-rich liquid phase that is in equilibrium with sulfur vapor (11). Consequently, it is not possible to synthesize $MoS_2$ via melt growth techniques and, instead, vapor phase methods are employed both for bulk crystals and thin films, employing sulfur-rich growth conditions where, according to the phase diagram, $MoS_2$ is in equilibrium with sulfur vapor.

The synthesis of TMD bulk crystals and thin films is dependent on the vapor phase supply of the chalcogen and transition metal to the growth front. Candidate sources for vapor phase synthesis of TMDs along with their respective melting temperatures ($T_{Melt}$) and vapor pressures ($P_v$) are given in **Table 1**. Vapor phase transport of chalcogens can be achieved in a straightforward manner via evaporation by heating elemental powders above their respective melting points. The transition metals, however, present a more significant challenge. Electron beam irradiation in a high vacuum environment is typically required for evaporation of Mo, W, etc. making this a suitable approach only for MBE. As a result, chemical precursors such as $MoO_3$, $MoCl_5$ and $Mo(CO)_6$ are employed. $MoO_3$, used in combination with S powder, is a popular source for powder vapor transport (PVT) growth of $MoS_2$, also referred to as powder source CVD. $MoO_3$ is air-stable and non-hazardous, however, it has a moderately high melting temperature (~802°C) and must be heated to >600°C to achieve an appreciable vapor pressure (12). In the PVT process, the $MoO_3$ powder is typically placed in close proximity to the substrate to achieve a sufficient vapor phase flux for growth (13).

Table 1. Physical properties of TMD source materials.

| Source | $T_{Melt}$ (°C) | Temp (°C) for $P_V$=1 Pa | $P_v$(Torr) at 25°C | Reference |
|---|---|---|---|---|
| Mo | 2622 | 2469 | $<10^{-6}$ | (15) |
| $MoO_3$ | 802 | 623 | $<10^{-6}$ | (12) |
| $MoBr_3$ | 500 | - | - | - |
| $MoCl_5$ | 194 | 60 | $<10^{-4}$ | (16) |
| $Mo(CO)_6$ | 150 | -1 | 0.177 | (17) |
| W | 3414 | 3204 | $<10^{-6}$ | (15) |
| $WO_3$ | 1473 | 1101 | $<10^{-6}$ | (12) |
| $WCl_6$ | 275 | 61 | $3.83\times10^{-4}$ | (18) |
| $W(CO)_6$ | 150 | 14 | 0.0415 | (19) |
| $WF_6$ | 2.3 | -101 | 1232 | (12) |
| Nb | 2477 | 2669 | $<10^{-6}$ | (15) |
| $NbCl_5$ | 205 | 52 | $4.5\times10^{-4}$ | (20) |
| Re | 3185 | 3030 | $<10^{-6}$ | (15) |
| $ReO_3$ | 400 | - | - | - |
| $ReF_6$ | 18.8 | -116 | 590 | (21) |
| S | 115 | 102 | $<10^{-5}$ | (10) |
| $(C_2H_5)_2S$ | -103.8 | -81 | 38.4 | (12) |
| $H_2S$ | -82 | -172 | $>10^4$ | (12) |
| Se | 220 | 216 | $<10^{-6}$ | (22) |
| $SeCl_4$ | 305 | 21 | 0.014 | (23) |
| $(CH_3)_2Se$ | -87.2 | -108 | 260 | (14) |
| $H_2Se$ | -65.7 | -145 | $>10^4$ | (12) |
| Te | 450 | 240 | $<10^{-6}$ | (22) |
| $(C_2H_5)_2Te$ | -10 | -66 | 10.3 | (24) |

Transition metal halides such as $MoCl_5$ and $MoBr_3$ have higher volatilities which translates into higher growth rates and therefore are better suited for bulk crystal growth using a chemical vapor transport (CVT) process. Halide compounds such as $MoCl_5$, however, are air sensitive and will react readily with ambient $H_2O$ to generate HCl, consequently, safety precautions are required. Metal carbonyls such as $Mo(CO)_6$ have moderate vapor pressures at room temperature and are ideal for use in epitaxial growth techniques such as MOCVD and metalorganic MBE (MOMBE), also referred to as hybrid MBE. However, these sources are toxic and will generate CO upon reaction and therefore are only suitable for use in a sealed source container. A similar trend in volatility is observed for W, Nb and Re sources (**Table 1**) with halide sources exhibiting the highest vapor pressure followed by the oxides.

Organo-chalcogen and hydride precursors are commonly used in place of elemental sources in MOCVD to avoid the requirements for source heating. Dimethyl selenium ($(CH_3)_2Se$) and diethyl sulfide ($(C_2H_5)_2S$), for example, have relatively high vapor pressures at room temperature (12, 14) although carbon contamination is a concern. The hydrides including $H_2S$ and $H_2Se$ are compressed gases but are toxic and corrosive and require special safety precautions.

**2.2 Chemical vapor deposition**

Powder vapor transport (PVT), or powder source CVD, is a common method for deposition of $MoS_2$ and related TMDs. In this process (**Figure 3a**), sulfur powder is placed in a boat and maintained at ~200°C; $MoO_3$ powder and the substrate are placed downstream and heated to 700-900°C. The furnace is slowly heated under a carrier gas to initiate vapor phase transport and growth. In general, TMD growth is limited by the supply of the transition metal precursor to the substrate surface, consequently, the nucleation density, growth rate and morphology of the TMD film are strongly dependent on the position of the substrate relative to the transition metal source (13). The S/Mo vapor phase ratio varies as a function of position in the PVT tube which alters growth stoichiometry, the relative growth rate of metal versus chalcogen terminated edges and the domain shape (25). PVT growth chemistry can also be altered through the addition of NaCl or KCl in the growth chamber. The addition of NaCl to the $MoO_3$ source results in the formation of $MoO_xCl_y$ species which have a higher volatility than $MoO_3$ (26). As a result, the flux of the Mo species is enhanced in the growth process resulting in the formation of large isolated domains that can be up to a millimeter in size.

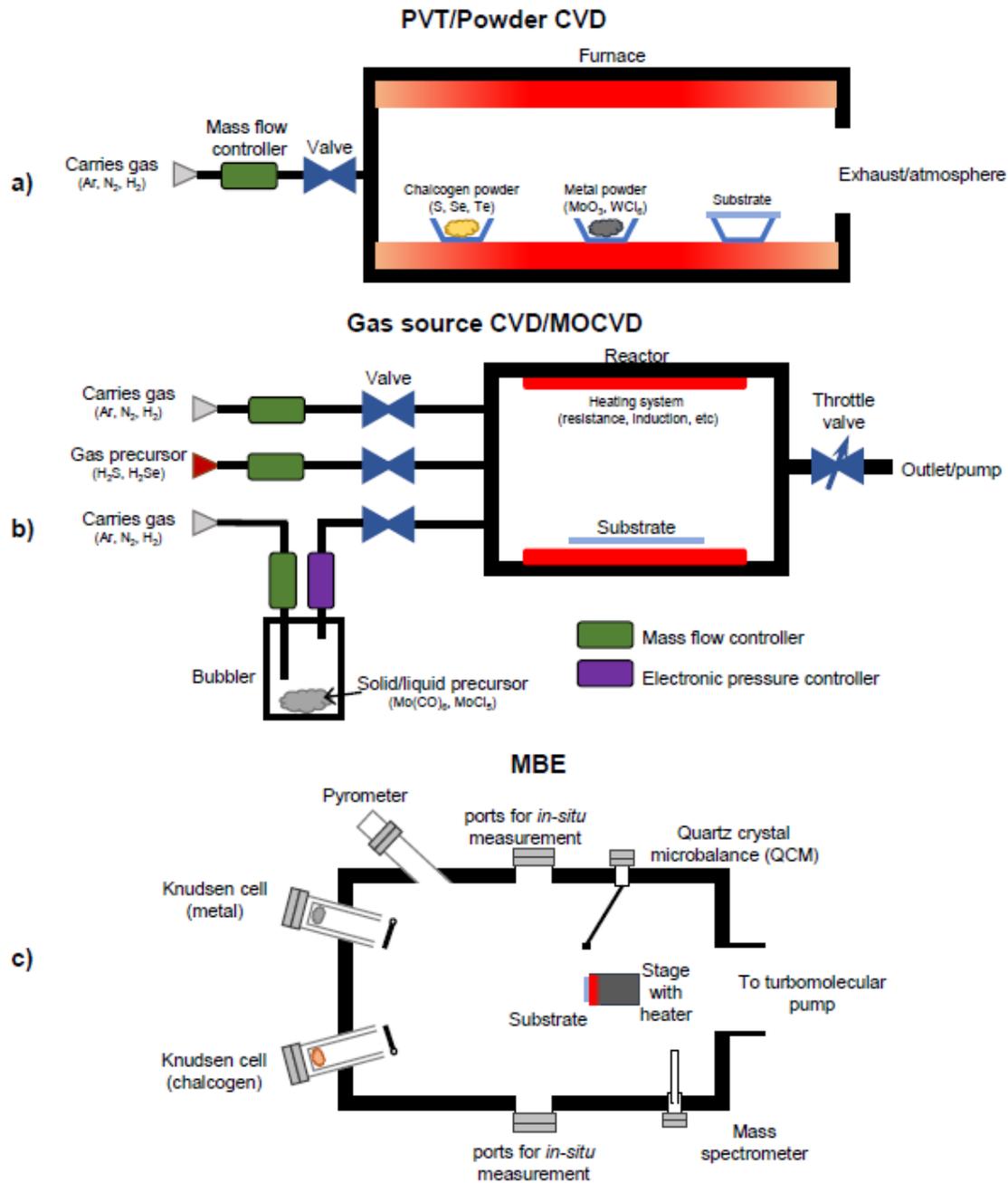

**Figure 3.** Schematic drawings of vapor phase deposition techniques for TMDs showing general source delivery and substrate heating configurations: (a) PVT/powder CVD; (b) Gas source CVD/MOCVD; (c) MBE.

While PVT is straightforward to implement, the source concentrations cannot easily be independently controlled and modulated which presents challenges and limitations. These difficulties can be overcome by using gas source CVD or MOCVD which utilizes volatile chemical precursors that are located outside of the deposition chamber (**Figure 3b**). MOCVD has previously been employed to synthesize a variety of TMD thin films (27-30). More recently, there has been renewed interest in MOCVD for the synthesis of monolayer and few-layer TMDs such as $MoS_2$ (31-33) and $WSe_2$ (34-36). A variety of precursors have been employed including metal carbonyls ($W(CO)_6$ and $Mo(CO)_6$), halides ($MoCl_5$, $WCl_6$, $NbCl_5$), organo-chalcogen compounds (($CH_3)_2Se$, $(C_2H_5)_2Se$, $(C_4H_9)_2Se$, $(C_2H_5)_2S$, $(C_4H_9)_2S$, $(C_2H_5)_2Te$, $(C_3H_7)_2Te$), and hydrides ($H_2S$, $H_2Se$). In the case of the organo-chalcogens, carbon deposition can occur at high temperature and high chalcogen/metal precursor ratios along with the TMD film (35, 37). With the exception of the hydrides, which are supplied as gases, the precursors are typically liquids or solids with low to moderate vapor pressure and are contained in temperature and pressure-controlled bubbler manifolds which employ a carrier gas for transport to the growth chamber.

## 2.3 Molecular Beam Epitaxy

Molecular beam epitaxy (MBE) is traditionally well suited for epitaxial growth of ultra-thin films since it offers a high level of control over source flux and the use of *in situ* characterization methods such as reflection high energy electron diffraction (RHEED), however, TMDs present special challenges. In a typical solid source MBE system (**Figure 3c**), effusion cells are used for evaporation of metals, however, the high melting temperature of transition metals such as Mo and W necessitates the use of electron-beam evaporation. In addition, the sticking coefficient of S (and Se to some extent) on the growth surface is low under ultra-high vacuum conditions impacting growth stoichiometry particularly at high growth temperatures which are beneficial to promote metal surface diffusion. As a result, TMD domain sizes are typically smaller than those achieved using PVT or MOCVD. To alleviate this problem, MBE growth of TMDs is often carried out on van der Waals materials such as epitaxial graphene, highly oriented pyrolytic graphite (HOPG) and mica, which enable enhanced surface diffusion (38). The growth of Te-based TMDs (e.g. $MoTe_2$ and $WTe_2$) which is typically carried out at lower temperatures than Se or S-based TMDs, is well suited for MBE growth. In addition, hybrid techniques such as MOMBE can be used employing $Mo(CO)_6$ or $W(CO)_6$ which eliminates the need for e-beam evaporation.

3. **Epitaxial Growth**

In epitaxial growth, the crystal symmetry of the substrate is used to control the crystallographic orientation of the film to ideally obtain single crystal (single domain) films with minimal dislocations and other defects. Layered materials present a unique situation for epitaxy due to the lack of out-of-plane chemical bonding, nevertheless, film-substrate interactions are sufficiently strong to cause in-plane alignment resulting in "epitaxy" despite lattice mismatch. This section includes an introduction to epitaxy, substrates considerations for TMDs and a review of recent literature for TMD epitaxy on traditional, metallic and van der Waals substrates.

## 3.1 Types of Epitaxy

Epitaxy can be categorized as conventional or van der Waals based on the substrate-film interaction which determines the impact of lattice parameter matching (**Figure 4**). In conventional epitaxy, strong bonding between the film and substrate makes lattice matching a crucial criterion for substrate selection. In such a scenario, a lattice mismatched film is strained until a critical thickness is reached above which the film relaxes by introducing misfit dislocations (**Figure 4a**). In van der Waals epitaxy, the substrate and the film are both layered materials and adhere via van der Waals forces (**Figure 4b**). Strict lattice matching rules are therefore relaxed but epitaxial films with sharp interfaces can still be achieved. This was first demonstrated in 1984, by Koma et al. for MBE growth of strain-free Se on cleaved Te and $NbSe_2$ deposition on $MoS_2$ despite lattice mismatches of 20% and 10%, respectively (39). Expanding on this concept, Koma et al. investigated passivating conventional 3D substrates for growth of van der Waals materials (**Figure**

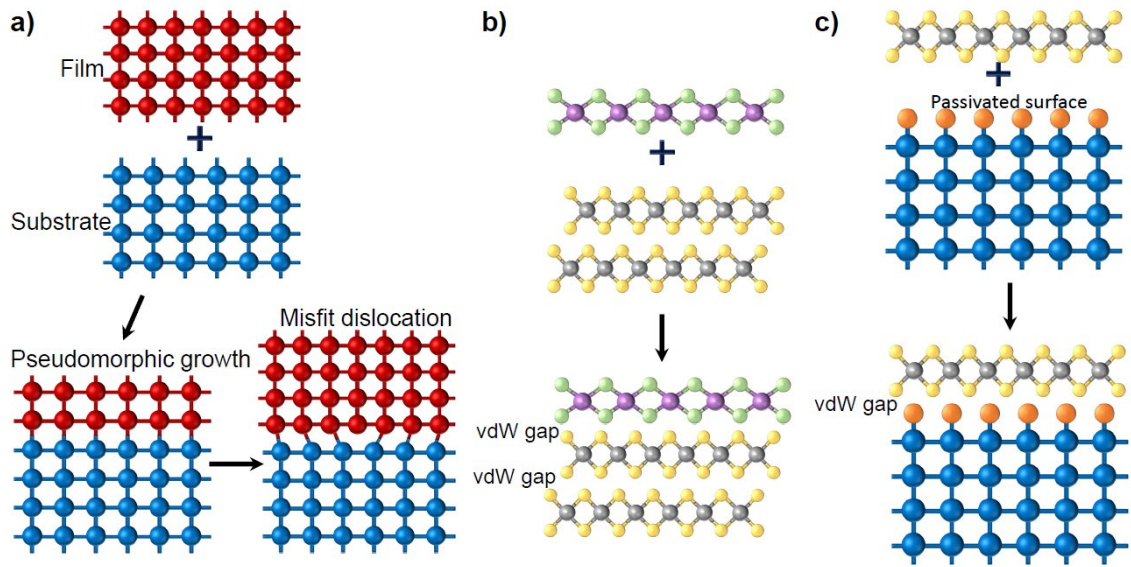

**Figure 4.** Schematic illustrations of (a) Conventional epitaxy where misfit dislocations nucleate to mitigate the stress due to lattice mismatch; (b) van der Waals epitaxy where both the films and the substrate have no dangling bonds and (c) van der Waals epitaxy on a passivated conventional 3D substrate.

**4c**) (40). For instance, epitaxial MoSe$_2$ was grown on hexagonal CaF$_2$ (111) face passivated by F. In Bi$_2$Se$_3$ growth on AlN (0001), misfit dislocations were observed in cross-sectional TEM, highlighting the importance of surface passivation (41). Silicon (111) surface terminated by Se (42) or reconstructed by Bi (43, 44) form an ideal van der Waals surface. Similar surface modifications also succeeded for the Al$_2$O$_3$ (0001) surface (45).

**3.2 Substrates**

Similar to thin film deposition of other materials, the substrate properties control the crystallinity of TMD films deposited by vapor phase methods. While polycrystalline TMD films can be obtained on amorphous substrates such as SiO$_2$/Si, epitaxial growth is required to obtain single crystal films. The substrate properties including crystal symmetry, lattice constant, miscut angle and surface energy are important factors to control the film orientation and surface coverage. There are a variety of candidate substrates with hexagonal symmetry that can be considered for TMD epitaxy (**Figure 5a**). Substrates such as (0001) sapphire ($\alpha$-Al$_2$O$_3$) and (0001) 4H or 6H SiC are suitable choices given their hexagonal symmetry, size, availability and good thermal and chemical stability. Van der Waals bonded materials such as graphene and hexagonal boron nitride (hBN) are also crystallographically compatible but are more difficult to obtain in large area. In addition, the low surface energy of graphene and hBN pose challenges to adsorption and nucleation. Wurtzite GaN is also interesting given the near lattice match with MoS$_2$ and WS$_2$.

In highly lattice mismatched film-substrate systems, superstructures can form to reduce the effective mismatch and promote epitaxy. If the film and substrate have similar symmetry but lattice parameters which differ by integral multiples of each other, a commensurate superstructure can form. For example, three unit cells of MoS$_2$ ($a$=3.160 Å) are commensurate with two unit cells of (0001) sapphire ($\alpha$-Al$_2$O$_3$) giving rise to an effective mismatch of ~0.4 % (**Figure 5b**) (46). The film may also rotate relative to the substrate such that certain lattice points of the film coincide with lattice points of the substrate. In this case, a superstructure referred to as a coincidence site lattice or Moiré pattern forms. This has been demonstrated in TMD epitaxy by MBE where MoS₂ rotated by 30° with respect to the (0001) sapphire reducing the in-place lattice mismatch to 13% since $\sqrt{3}a(MoS_2)$=5.470 Å (47).

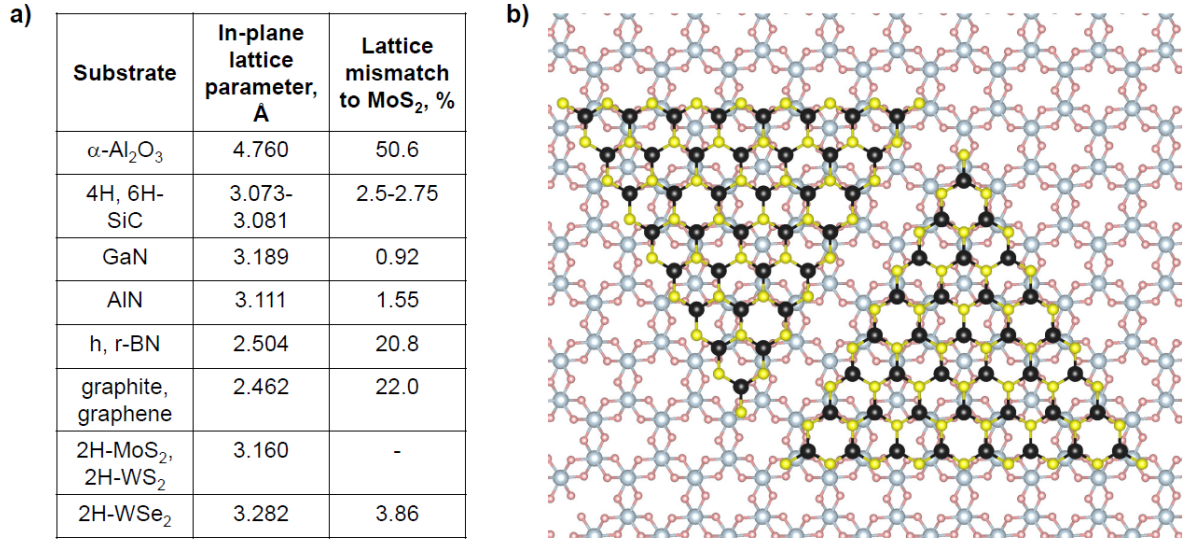

**Figure 5.** (a) Comparison of in-plane lattice parameters and lattice mismatch to MoS₂ for candidate substrates. (b) Schematic of MoS₂ on sapphire showing 3(MoS₂)x2(α-Al₂O₃) commensurate structure and MoS₂ domains oriented at 0° and 60° (figure prepared using VESTA http://jp-minerals.org/vesta/en/).

In the case of TMDs, due to the inherent symmetry, there is no energetic difference between domains oriented at 0° and 60° with respect to each other (**Figure 5b**) on high symmetry substrates. Due to the broken inversion symmetry in TMDs, when domains of opposite orientation coalesce, an anti-phase boundary (APB) forms, also referred to as an inversion grain boundary. The APBs have a metallic character which is detrimental to electronic properties (48-51) and is also problematic for studies of valley-related physics. TMD domains should ideally exhibit the same orientation and alignment thereby enabling them to seamlessly coalesce. The formation of anti-phase domains in TMD films can be reduced, to some extent, through substrate modification. Control of the substrate miscut and step height has previously been used to reduce APBs in epitaxial growth of polar films on non-polar substrates such as III-Vs on Si (52). For layered materials, the presence of surface steps has been shown to reduce twinning in $Bi_2Se_3$ films grown by MBE on (111)Si (43) and to control domain orientation in $WSe_2$ (53) and $MoSe_2$ (54) grown by PVT on (0001) sapphire. In the case of van der Waals substrates, substrate defects may provide a pathway to break the surface symmetry and promote single orientation films, as demonstrated for $MoS_2$ (55) and $WSe_2$ (56) on hBN.

### 3.2.1 Conventional Substrates

Sapphire is a commonly used insulating substrate due to its large size at relatively low cost and good chemical and thermal stability at high temperature and under various growth environments. C-plane (0001) sapphire has been the most widely used for TMD epitaxy due to the hexagonal symmetry of the surface, (46, 47, 58-60) although TMD growth on other oriented surfaces of sapphire such as r-plane ($1\bar{1}02$) has also been studied. Dumcenco et al. first demonstrated two major equivalent orientations at 0º and 60º for CVD-grown $MoS_2$ domains on c-plane sapphire as well as commensurability between the lattices (**Figure 6 a,b**) (46). Similar results have been observed in fully coalesced centimeter-scale $WSe_2$ monolayers grown by MOCVD on sapphire, although the films exhibit non-uniform optical and electrical properties owing to the strong coupling between the TMD film and sapphire surface steps and the strain from thermal expansion mismatch which occurs during cooldown from the growth temperature.(58, 60) Preferential nucleation and orientation of TMD domains along $[11\bar{2}0]$ step-edges on sapphire has also been reported, offering the possibility to reduce APBs in coalesced films. Chen et al. reported step-edge alignment of $WSe_2$ on sapphire in PVT growth at higher temperatures (950ºC) (53). Hwang, et al. observed similar nucleation of $MoSe_2$ at sapphire step-edges (**Figure 6c**) which was dependent on the PVT carrier gas, flow rate and growth temperature (54).

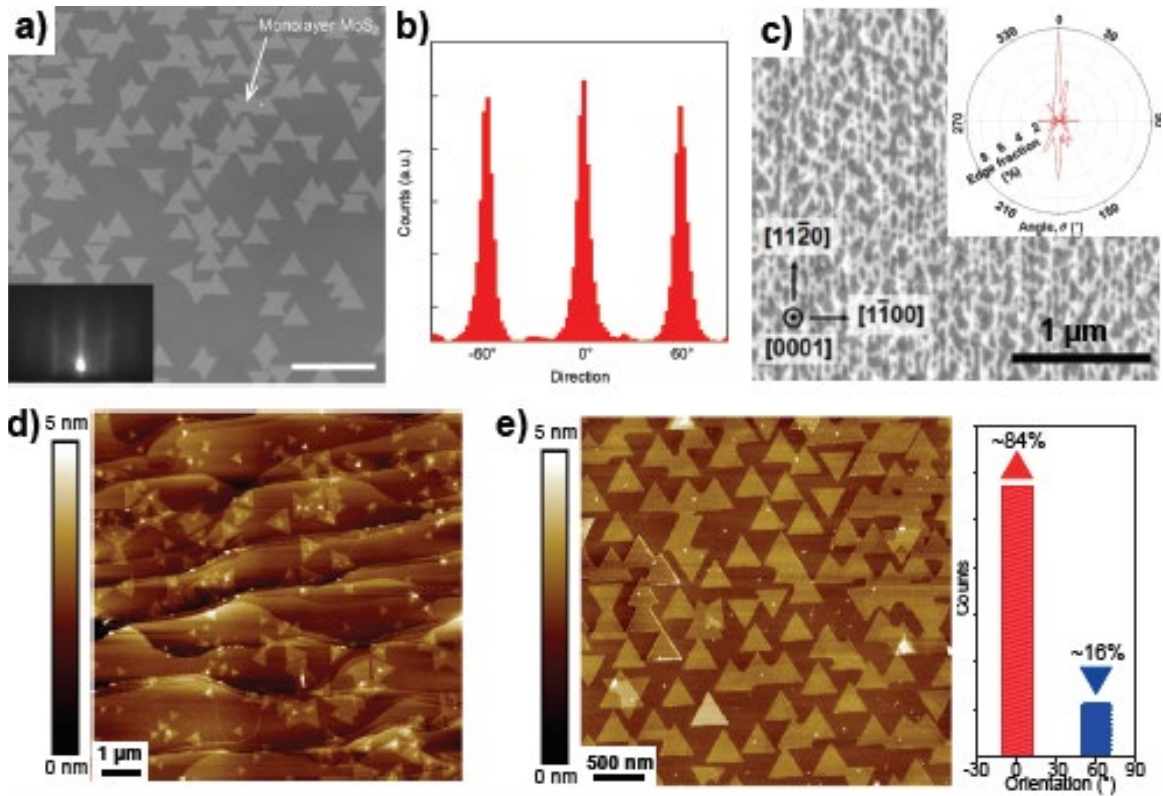

**Figure 6**. (a) Optical microscopy image of monolayer $MoS_2$ grains grown on c-plane sapphire. Inset: RHEED pattern acquired on the sample showing the film with long-range structural order. Reprinted with permission from (46). (b) Orientation histogram based on the area shown in part a confirms that the majority of $MoS_2$ grain edges are oriented along 0º and 60º angles. Reprinted with permission from (46). (c) SEM image of $MoSe_2$ aligned along step-edges for growth temperature of 750 °C. Inset: Polar plots of the $MoSe_2$ edge length fractions oriented with the angle, θ from [11$\bar{2}$0] substrate orientation. Reprinted with permission from (54). (d) AFM image of $WS_2$ domains nucleated on defect sites cleated by plasma treatment of EG. (e) AFM image of epitaxial $WSe_2$ domains on exfoliated hBN flakes showing a ~84 % single orientation. Reprinted with permission from (56).

The epitaxial growth of TMDs on III-V substrates is also of interest for hybrid 2D/3D electronics and optoelectronic applications. Ruzmetov et al. reported the growth of monolayer $MoS_2$ domains on (0001) GaN, where the edges of $MoS_2$ align with the m-plane (1$\bar{1}$00) of GaN (61). The lattice constants and thermal expansion coefficients of $MoS_2$ and GaN are similar, therefore the MoS$_2$ was unstrained. Similar results were reported by Wan et al., who also reported TMD epitaxy only on n-type GaN that had a lower roughness compared to p-type GaN (62). In addition, monolayer MoSe$_2$ has been epitaxially grown by MBE on GaAs(111)B, resulting in an epitaxial relationship of MoSe$_2$[10$\bar{1}$0]//GaAs[11$\bar{1}$] and MoSe$_2$[11$\bar{1}$0]// GaAs[10$\bar{1}$] (63).

Additional substrates have been explored for TMD epitaxy. For example, the formation of well-aligned vertical MoS$_2$ "fins" on 4H- and 6H-SiC substrates was demonstrated by PVT for catalytic applications (64, 65). In this case, the epitaxial relationship is MoS$_2$[0001]//SiC<11$\bar{2}$0> and arises due to the high MoO$_3$:S ratio used for growth compared to a lower ratio which resulted in planar domains. Single crystal complex oxides can also be a potential substrate for TMD epitaxy. Chen et al. discussed the epitaxial growth of monolayer MoS$_2$ on (111), (110), and (001) terminations of single crystal SrTiO$_3$ (66). Although epitaxial growth of MoS$_2$ can be realized on all three surfaces, the domain morphology and size exhibit with distinct differences indicating the surface energy can still strongly regulate the epitaxy of 2D TMDs.

Crystalline metals have also been used as substrates for epitaxial growth of TMDs, for example, large domains of monolayer TMDs have been demonstrated on Au (111) substrates (67-70). Bana et al. demonstrated commensurability of (10×10)MoS$_2$ over (11×11)Au(111) and reported that a single domain orientation of MoS$_2$ was achieved enhancing spin-polarization of the monolayer (70). Meanwhile, Shi et al. found that large MoS$_2$ domains are more preferentially evolved on Au(100) and Au(110) facets than on Au(111) due to different binding energies of MoS$_2$-related species on the facets (71).

### 3.2.2 van der Waals Substrates

Van der Waals substrates like mica, graphene and hBN have been explored for the growth of TMDs. Koma et al. demonstrated strain-free $NbSe_2$ and $MoSe_2$ on mica using MBE (72). $MoS_2$ (73) $ReS_2$ (74,75) and $VSe_2$ (76) have also been deposited on mica by PVT, and exhibit the presence of mirror domains. As mica undergoes dehydroxylation at ~700°C (77) excursions at higher temperatures can have significant impact on growth (74).

In case of graphene, TMD growth has been investigated on HOPG, epitaxial graphene on SiC (EG) and graphene synthesized by CVD on copper. HOPG steps act as nucleation sites where the TMD domain orientation is influenced by the step-edge termination (78, 79). EG, synthesized by sublimation of SiC, has tunable electronic properties determined by the interaction between the carbon layers and the underlying SiC. As-grown EG has a covalently bonded buffer layer resulting in strain of 1% in the top layers (80). Hydrogenation passivates the SiC surface and releases the buffer layer and the inherent strain, resulting in quasi free-standing graphene (QFEG) (81). Hydrogenation can also cause wrinkles due to thermal mismatch (82). It was observed that $MoS_2$ thickness was higher on as-grown EG than QFEG (83). The difference can be attributed to the graphene thickness and the strain which affects the chemical reactivity of graphene. Defects created on EG can also increase nucleation by increasing the chemical reactivity, as shown in **Figure 6d**. In addition, wrinkles and steps edges present on EG and QFEG significantly influence the growth (83). Similar effects have been observed for CVD graphene transferred on to a different substrate (84) or suspended on a TEM grid (85). The quality of the graphene and the transfer process control the nucleation sites. Growth is rarely done without transfer due to the underlying copper substrate. Low temperature $MoS_2$ growth on graphene on copper, shows the presence of multiple grains with orientations distributed in a narrow range -11 to 18° with respect to a single grain in the graphene film (86). In all types of graphene surfaces, mirror domains are observed, which on coalescence will lead to APBs.

Hexagonal boron nitride (hBN) provides a promising alternative substrate. hBN has previously been used as a dielectric layer and an encapsulant for protecting TMDs and enhancing its properties (87-89). TMD growth on hBN has been attempted using BN films grown by CVD and exfoliated flakes. CVD BN grown on sapphire or on copper and transferred to a $SiO_2$/Si substrate show that the TMD film quality is affected by the quality of the underlying BN layers (90). Multiple layers of BN work more efficiently, as BN degrades in $H_2$ or chalcogen precursor flux at high temperature (90). Growth of TMDs on exfoliated BN showed an improved film quality (91-93). In some cases, mirror domains were found to be equally probable (92). Fu et al., observed that a single orientation was favored in MBE when lower metal flux was used (91). Similar single oriented domains have been observed for $WS_2$(95) and $WSe_2$ (**Figure 6e**) (56). Preference of a single orientation was explained based on metal atoms reacting with the BN defects and acting as nucleation or anchoring sites. The presence of these anchoring sites removes the degeneracy favoring one orientation preferentially (55). In addition, Zhang et al. demonstrated control over the nucleation density using a combination of plasma treatment and $NH_3$ annealing (56). However, large area films would need the synthesis of large area high quality hBN substrates.

## 4. Characterization of epitaxial TMD monolayer films

In epitaxially grown TMDs, it is imperative to characterize layer thickness, the degree of crystallinity, the nature of grain boundaries as well as strain due to epitaxy which govern the global optical and electrical properties of the films. In the following sections, an emphasis on characterization approaches for layers directly on the growth substrate is discussed with examples for TMD films grown epitaxially on crystalline substrates.

## 4.1 Identifying layer number

Spectroscopic techniques such as photoluminescence and Raman spectroscopy have been conventional probes of layer number in 2D materials. Scanning probe techniques, such as atomic forces microscopy (AFM) also allows for quick determination of layer number and layer homogeneity in large area TMD films directly on the growth substrates. Other approaches include low-energy electron microscopy (LEEM), which is a well-known method to characterize epitaxial graphene. The short penetration and escape depth of the incident electrons in LEEM provides high spatial surface sensitivity of the structural and electronic properties of grown films and thus is uniquely suited to study epitaxial TMD monolayers and heterostructures. Through a subset measurement in LEEM known as low-energy electron reflectivity (LEER), de la Barrera et al. showed that it is possible to extract information on the number of layers in epitaxially grown $WSe_2$ on epitaxial graphene by measuring the spectral intensity of reflected electrons as a function of beam energy (95). From electron reflectivity maps shown in **Figure 7 a,b,** spectra related to the electronic localized states between individual layers can be extracted and thus correlated to layer thickness extracted from other probes, such as AFM and Raman spectroscopy. Other surface sensitive measurements techniques such as spectroscopic ellipsometry (SE) have been useful in characterizing epitaxial TMD films. Eichfeld et al. extracted the refractive index (n) and extinction coefficient (k) with SE in MOCVD grown $WSe_2$ films on sapphire with different layer thicknesses (96). Through SE, film thickness was determined within a 9% accuracy to measurements performed by AFM.

## 4.2 Epitaxial relationship and crystallinity

X-ray diffraction (XRD) has been widely implemented for the characterization of solid materials to determine the degree of crystallinity, crystal structure parameters and orientation of

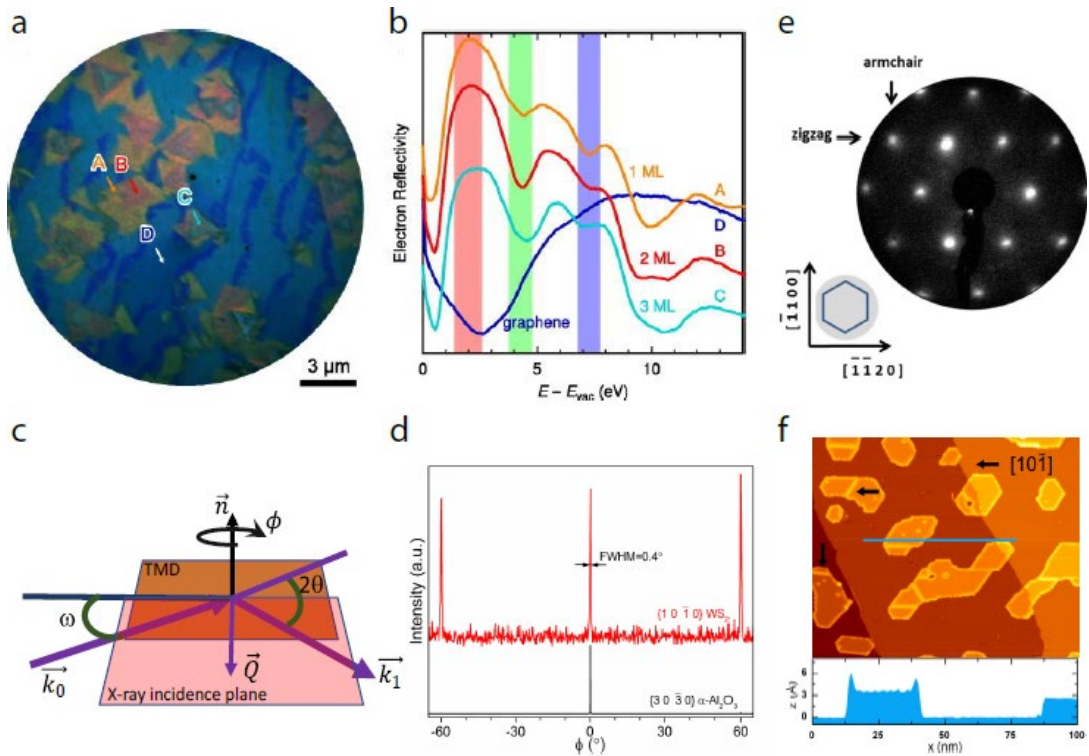

**Figure 7. Characterization of epitaxial TMD monolayer films.** LEER characterization of layer number: (a) Spectroscopic image of MOCVD-grown WSe$_2$ on epitaxial graphene with colors assigned to the intensity of reflected electrons at specified energy slice shown in (b). Reprinted with permission from (95). In-plane XRD of monolayer TMD: (c) XRD setup showing TMD surface with its surface normal n, X-ray incidence plane defined by incident and diffracted wave vectors k with 2θ angle between them and scattering vector Q as well as instrument angles ω and $\phi$. (d) Upper red curve - $\phi$-scan of {1 0 $\bar{1}$ 0} planes of WS$_2$ and bottom black curve – $\phi$ -scan of {3 0 $\bar{3}$0} planes of α-Al$_2$O$_3$. (e) LEED pattern illustrating lattice alignment between grown MoS$_2$ and the sapphire substrate. Reprinted with permission from (99). (f) STM image highlighting the 1D metallic nature of the mirror twin grain boundaries in MoS$_2$ grown on graphene/Ir(1 1 1) by MBE. Reprinted with permission from (108).

crystallites. Implementation of XRD in TMD films would be beneficial as this technique does not require sample preparation, it is nondestructive and provides information over large areas (~cm$^2$). To study monolayer 2D materials with XRD, it is important to measure planes that are perpendicular to the layer. In graphene, due to the low x-ray scattering factor of carbon, it is challenging to study even multilayers films using XRD instrumentation that employs static anode

Cu X-ray tubes. However, the high x-ray scattering factor of transition metals and chalcogens makes it possible for standard diffractometers to perform XRD measurements on monolayer TMD films, thus eliminating the need for synchrotron-based techniques. Laboratory scale XRD instruments that use static anode Cu X-ray tubes with parallel beam optics can used to determine monolayer TMD in-plane crystal orientation and estimation of the crystalline quality by observing twist of the crystal (97). In this setup, the x-ray incidence plane is ~0.5 - 1° above the surface of the TMD film. Initial alignment is performed on {h k 0} planes of the substrate utilized. After that, setting 2θ at {h k 0} planes of TMD and ϕ-scan (sample rotation around surface normal) can be performed. This measurement allows one to determine whether the TMD crystal has one or multiple in-plane orientations and its degree of alignment to the substrate by comparing ϕ angles of the peaks of {h k 0} planes of the film to those of the substrate. **Figure 7 c,d** illustrates the instrument setup and ϕ-scan of $WS_2$ {10$\bar{1}$0} planes plotted along with {30$\bar{3}$0} related peaks of sapphire, showing the epitaxial orientation. In addition, crystal quality of the TMD materials can be assessed by measuring the full width at half maximum (FWHM) of the ϕ peak which indicates the in-plane twist of the crystal.

### 4.3 Domain size and orientation

New emergent functionalities seen in exfoliated 2D monolayers from bulk crystals, have made characterizing domain orientation in synthetic large area grown films a topic of important consideration. Particularly in spintronic and valleytronic devices fabricated from single TMD monolayers, the broken inversion symmetry and strong spin-orbit interactions gives rise to distinct spin and valley-dependent optical selection rules. For example, transitions between the K and −K valleys can dependend on the handedness of circularly polarized light (98). However, demonstrations of these emergent degrees of freedom have only been realized in devices fabricated from exfoliated bulk single crystals. Reports using wafer scale CVD and MBE grown TMD monolayer films are seldomly discussed in the literature. Single domain-orientation is an essential requirement in large area coalesced films to clearly distinguish transitions that occur at the K and −K points in the Brillion zone. The associated bandstructures of two degenerate mirror domains in monolayer TMD films leads to spin reversal in the valence band extrema at the K and −K points. Mirror-symmetric domains with a degenerate 0° and 60° orientation are commonly observed in epitaxial TMD films grown on sapphire, GaN, etc. and their density can impact the observation of spin and valley polarization. Therefore, it is imperative to characterize domain orientations with respect to the substrate, their size and density in the coalesced film.

The lattice orientation of grown films can be characterized using low-energy electron diffraction (LEED) in LEEM. Yu et al. demonstrated the lattice alignment of CVD grown $MoS_2$ monolayer on sapphire with LEED (100). In **Figure 7e**, from the sharp hexagonal diffraction LEED pattern of $MoS_2$ on sapphire, no changes in the LEED diffraction spots were further observed when the sample was tilted in orthogonal directions. This suggests preferential epitaxial relationship of the zigzag and armchair directions of $MoS_2$ to the $[\bar{1}\bar{1}20]$ and $[\bar{1}100]$ directions of sapphire, respectively. However, it is important to note that LEED measurements cannot distinguish the difference between the two degenerate 0° and 60° TMD domain orientations that can commonly coexist when epitaxial growth is performed on sapphire substrates. To determine domain orientation, $MoS_2$ films were transferred onto a carbon grid for high resolution transmission electron microscopy (HR-TEM) and a domain orientation of 0° and 60° was identified by observing the existence of the mirror-twin grain boundary in the coalesced films.

Domain orientation can be determined in x-ray photoelectron diffraction (XPD), which is based on emission-angle dependent photoemission intensity modulations of core levels in different layers of atoms (100-102). Intensity modulations from XPD measurements have been successfully used in studying domain orientation in hBN and $MoS_2$. These intensity modulations reflect the local structure of the emitting surface atoms. By analyzing the geometric structure in the stereographic projections of modulation in the core level intensities (correlated with XPD simulations), the amount of mirror orientation of domains in coalesced layered can be extracted. Moreover, spin-resolved photoemission studies, such as spin-polarized angle resolved photoemission spectroscopy measurements have also proven useful in determining the influence of the degree of mirror domain orientation on the bandstructure in TMD films (100, 103).

### 4.4 Nature of grain boundary in TMD monolayer films:

When two mirror domains coalesce, a special low symmetry grain boundary develops between them known as a mirror-twin grain boundary. In monolayer films, these grain boundaries are 1D line defects that are metallic in nature, can undergo charge density wave (CDW) transitions and are 1D strongly correlated quantum liquids (104). Different types of mirror-domain boundaries can exist that differ in structure and formation energy. Formation of mirror-twin grain boundary loops is prominently observed in MBE samples grown under an excess of transition metal, examples in transition metal di- selenides and tellurides grown films under transition metal-rich conditions are widely reported (105, 106). Understanding the formation of these grain boundaries, their structure and electronic properties is of critical importance in large area grown films. As alluded to previously, through post growth transfer, mirror-twin grain boundaries in TMD monolayers can be investigated in HR-TEM. These boundaries commonly observed in HR-TEM are chalcogen deficient (107). Scanning tunneling microscopy (STM) has also proven useful in characterizing the formation of such grain boundaries in TMDs and their electronic structures. In $MoS_2$ grown on graphene/Ir(1 1 1), the grain boundaries appear bright and prevalent in STM (**Figure 7f**), which is indicative of their 1D metallic nature (108). It is possible to control the formation of these grain boundaries by breaking the degeneracy in the mirror domain formation. This has been demonstrated through careful growth of $MoS_2$ on Au (1 1 1) and, $MoS_2$ and $WSe_2$ grown on treated hBN (55, 56, 100). Like mirror-twin grain boundaries, edges in isolated domains from incomplete coalescence can also form. In these isolated domains, different edge termination also give rise to unique electronic and chemical properties.

## 5. Future Outlook

Significant progress has occurred in recent years in achieving wafer-scale epitaxial TMD monolayers, but further advances are needed. By controlling the substrate surface structure and defects, the formation of mirror twins has been reduced resulting in TMD domains that are nearly single orientation on both sapphire (54) and hBN (55, 56). Nevertheless, detailed characterization of grain boundaries and other defects in coalesced TMD epitaxial monolayers is generally lacking and additional work is needed to suppress bilayer formation across wafer-scale areas. This work would benefit from the availability of *in situ* characterization techniques operable at relevant growth temperatures and pressures that would enable measurements of domain size and orientation at the sub-monolayer level. While sapphire is the most common substrate used to date, thermal strain and substrate charge can negatively impact the uniformity of optical and transport properties of the TMD monolayer. Consequently, alternative substrates or buffer layers should be examined such as hBN which provides an ultra-smooth passivated surface and dielectric screening. In addition, non-destructive, full-wafer characterization methods are needed to measure surface coverage, defect density, optical properties and transport characteristics of monolayer TMD films.

Moving beyond monolayers, growth of lateral and vertical TMD heterostructures represents the next step in epitaxy development. This has been examined, to some extent, for isolated TMD domains grown by PVT or MOCVD but not in detail for wafer-scale films. Vertical heterostructures formed by epitaxy offer the prospect of achieving ultra-clean interfaces with strong electronic and optical coupling between layers. However, fundamental issues related to heteroepitaxial growth have not yet been examined. For example, in a heterostructure, the underlying monolayer serves at the "substrate" for the second layer and hence the properties of this layer including strain, point defects, grain boundaries, etc. will impact precursor adsorption, nucleation, wetting, lateral growth and coalescence. For mixed S/Se/Te vertical heterostructures, chalcogen exchange can occur depending on the growth chemistry and conditions employed. In the case of lateral heterostructures, opportunities exist to employ epitaxial growth to create quantum dots of a narrower gap material embedded in a wider gap matrix, lateral superlattices with anisotropic properties or artificially ordered alloys. Further developments will be required to control nucleation site and density and understand the impact of lattice mismatch and relaxation mechanisms along the edge-face of TMD crystals.

## Acknowledgements

The authors acknowledge the financial support of the National Science Foundation (NSF) through the Penn State 2D Crystal Consortium – Materials Innovation Platform (2DCC-MIP) under NSF cooperative agreement DMR-1539916.